\documentclass[twocolumn]{aastex631}

\shorttitle{Making the invisible visible}
\shortauthors{Barnier \& Done}

\graphicspath{{./}{}}

\usepackage{graphicx}	
\usepackage{amsmath}	
\usepackage{url}

\begin{document}

\title{Making the invisible visible: \\Magnetic fields in accretion flows revealed by X-ray polarization}

\author[0000-0002-4180-174X]{Samuel Barnier}
\affiliation{Theoretical Astrophysics, Department of Earth and Space Science, Graduate School of Science, \\ Osaka University, Toyonaka, Osaka 560-0043, Japan}
\email{sbarnier@astro-osaka.jp, chris.done@durham.ac.uk}

\author[0000-0002-1065-7239]{Chris Done}
\affiliation{Centre for Extragalactic Astronomy, Department of Physics, Durham University, South Road, Durham DH1 3LE, UK\\
Kavli Institute for the Physics and Mathematics of the Universe (WPI), University of Tokyo, Kashiwa, Chiba 277-8583, Japan}

\begin{abstract}

Large scale, strong magnetic fields are often evoked in black hole accretion flows, for jet launching in the low/hard state and to circumvent the thermal instability in the high/soft state. Here we show how these ideas are strongly challenged by X-ray polarization measurements from IXPE.  Quite general arguments show that equipartition large scale fields in the accretion flow should be of order $10^{6-8}$~G. These produce substantial Faraday rotation and/or depolarization. Since IXPE observes polarisation in both spectral states, this sets upper limits to coherent large scale (vertical, radial or azimulthal) magnetic fields in the photosphere of $B\lesssim 5\times10^6$~G. 
While we stress that Faraday rotation should be calculated for each individual simulation (density, field geometry and emissivity), it seems most likely that there are no equipartition strength large scale ordered fields inside the photosphere of the X-ray emitting gas. Strong poloidal fields can 
still power a Blandford-Znajek jet in the low/hard state if they thread the black hole horizon rather than the X-ray emitting flow, but this could also be challenged by (lack of) depolarisation from vacuum birefringence. Instead, an alternative solution is that the low/hard state jet is dominated by pairs so can be accelerated by lower fields. Strong toroidal fields could still stabilise the disc in the high/soft state if they are buried beneath the photosphere, though this seems unlikely due to magnetic buoyancy.
Fundamentally, polarization data from IXPE means that magnetic fields in black hole accretion flows are no longer invisible and unconstrained. 

\end{abstract}


\section{Introduction}

Current consensus is that jets from black hole accretion flows are powered by a combination of rotation and magnetic fields.
There are two main models for their formation, either using the rotational energy of the black hole (\citealt{BZ1977}, hereafter BZ, where the jet is spin powered), or the accretion flow (\citealt{BP1982}, hereafter BP, or accretion powered jet). These differ also in the magnetic field configuration required. The BZ process uses large scale vertical (poloidal) field, but threading the black hole horizon rather than the accretion flow, so the back reaction slows down the black hole spin.  The BP process again uses the vertical (poloidal) component of a large scale ordered field threading the accretion flow to accelerate a small fraction of the matter in the flow upwards into the jet. The back reaction of the torque onto the accretion flow slows down the accretion flow rotation, allowing it to fall inwards by transporting 
its angular momentum upwards with the outflow. 

The discovery of the magneto-rotational instability (MRI, e.g. \citealt{Velikhov1959,Chandrasekhar1960,Balbus1991,Hawley1995,Balbus1998,Sorathia2012,Hawley2013}) held out the hope that the magnetic field structures could be calculated from first principles. The MRI produces a small scale, turbulent magnetic dynamo, amplifying any weak field present in the flow. Decades of work have shown that the instability grows linearly, then saturates in the non-linear regime, giving a magnetic field structure with well defined average properties in the flow, forming both turbulent and ordered fields. 

The problem is that there has been  no way to observationally test the predicted magnetic field structures. Here, we show that this is now possible with the advent of X-ray polarimetry, and that current data from the new Imaging X-ray Polarimetry Explorer (IXPE) provides stringent observational constraint on the magnetic fields due to Faraday rotation. 

We demonstrate this using the IXPE data from the 
jet launching low/hard states of stellar mass black hole binary systems. These states show steady radio jets that probably require the presence of magnetic fields to be launched/collimated.
The correlation between the radio jet luminosity and the X-ray luminosity of the accretion flow observed in the low/hard state \citealt{Corbel2000,Corbel2013,Coriat2011}) also suggests a strong connection between the accretion and ejection processes.
Scattering imprints polarization onto the X-rays, and the integrated signal over the entire X-ray hot plasma has polarization 
fraction and angle which are diagnostics of the X-ray source geometry. The less spherically symmetric  the geometry, the more polarization imprinted.
The polarization angle then shows the seed photon direction relative to the  line of sight to the observer. 
A planar disc has polarization from electron scattering which increases
with inclination, and the angle switches from aligned parallel to the disc plane for optically thick emission (seed photons travelling vertically before scattering) to aligned perpendicular to the disc plane (parallel to the jet) for optically thin plasma (seed photons travelling horizontally in order to intercept an electron and be scattered). The hard X-rays seen in the jet emitting state in black hole binaries are 
from optically thinnish ($\tau\gtrsim1$), hot plasma, so the angle of polarization should be directed perpendicular to the plane of the material i.e. along the jet axis. 

The IXPE observations of Cyg X-1 in the jet emitting state show that the hard X-rays are polarized at a level of $\sim$5\%, and that the polarization angle is aligned with the radio jet as imaged on the sky (\citealt{Krawczynski2022}). This means that the X-ray plasma is extended perpendicular to the jet, consistent with a geometry where the hot plasma is a radially extended accretion flow. This amount of polarization is already quite large given that the inclination of the binary is only $\sim 30^\circ$. The simplest solution is that the inner accretion flow is misaligned with the binary axis, so that it is viewed at higher inclination. There is no change in expected polarization angle as we see only the projection of the plane of the accretion flow on the sky rather than its full 3 dimensional alignment. 

We show below how Faraday rotation of the polarization plane puts a stringent constraint on the vertical (poloidal) magnetic field inside the X-ray hot plasma in Cyg X-1 to $\lesssim 2\times10^6$~G.  
Other large scale ordered field components (radial and azimuthal) also have strict upper limits around $5\times10^6$~G as the Faraday rotation direction switches across the disc, leading to depolarization (see also \citealt{Gnedin2006}).
These fields are typically sub-equipartition strength with the gas pressure in the flow for the bright low/hard states, whereas BP jets typically have fields which are equipartition or larger. 
Jet launching 
is still possible at sub-equipartition fields 
($|B|\lesssim 10^6$~G, \citealt{Jacquemin2019}), but the resulting jet
powers are likely small.

We also discuss IXPE data from black hole binaries in their disc dominated soft states. The standard Shakura-Sunyaev disc becomes unstable when the total pressure inside the disc is dominated by radiation rather than gas pressure (\citealt{Lightman1974,Shakura1976}). 
This occurs for $L/L_{Edd}\gtrsim 0.06$   but the predicted limit cycle behaviour (\citealt{Szuszkiewicz1997,Szuszkiewicz1998,Honma1991a,Zampieri2001})
is not seen in the data e.g in LMC X-3 where stable discs are seen up to at least $L/L_{Edd}\sim 0.5$
(\citealt{Gierlinski2004,Steiner2010}). One way to supress the instability is if the disc has substantial magnetic pressure support (e.g. 
\citealt{Begelman2007,Oda2009,Sadowski2016,Mishra2022}). 
Large toroidal fields are especially likely as rotation will coherently wind up this component (e.g. \citealt{Blaes2006,Begelman2007,Bai2013,Begelman2024}). 
However, the observed polarizations from  optically thick (disc-like) spectra in 4U1630-40 and Cyg X-1 challenge this, as (lack of) Faraday rotation/depolarization limits $B_\phi \lesssim 5\times 10^6$~G which is far below equipartion.  

We conclude that the new IXPE data mean that magnetic fields are no longer a free parameter in any model. Polarization data mean there are now observational constraints, and that these are quite stringent.
Large scale ordered fields with 
strength around equipartition with the gas pressure in the flow are ruled out, and even fields separated from the flow (e.g. pinned onto the horizon) give observational signatures for strengths around equipartition with the ram pressure. This challenges multiple models of jet formation and disc stabilization. 

\section{polarization and Faraday rotation}
Polarization is a fundamental feature of electromagnetic waves, but we review it here as observational data are still very new in the X-ray 
waveband. 
A wave travelling along the z-axis in standard cartesian coordinates is 
completely linearly polarized if the electric field vector 
$\vec{E}$ has components $E_x(t)=E_x(0)\cos(\omega t-\Phi)$ and 
$E_y(t)=E_y(0)\cos(\omega t-\Phi)$. 
This defines a wave 
in the $x-y$ plane, making angle $\Phi=\arctan[E_y(0)/E_x(0)]$ to 
the $x$ axis (in this paper we ignore circular/elliptical polarization). 

\subsection{Adding linearly polarized waves}

The observed emission is the sum over multiple photons, so 
if all orientations are equally probable then the total emission is unpolarized. However, 
some processes such as electron scattering preferentially result in some polarization angles being more likely. The fraction of polarization $p$ in the total beam is defined as $p=I_p/I$ where $I_p$ is the intensity of the polarized beam and $I$ is the total intensity. However, adding over multiple photons is not straightforward as polarization is not linear. It is easiest to see this via the polarization angles. 
If $E_y(0)=0$ then the electric vector $\vec{E}$ is completely in 
the plane $y=0$, which is the same plane as for an angle $\Phi=\pi$. The orthoganal polarization component, 
is for $E_x(0)=0$ which is a rotation of $\pm\pi/2$. A beam which is made from equal amounts of plane waves with $E_x(0)=0$ and $E_y(0)=0$ is unpolarized, so we need a set of equations to combine polarizations which have this property.

\begin{figure}
    \centering
    \includegraphics[width=0.5\textwidth]{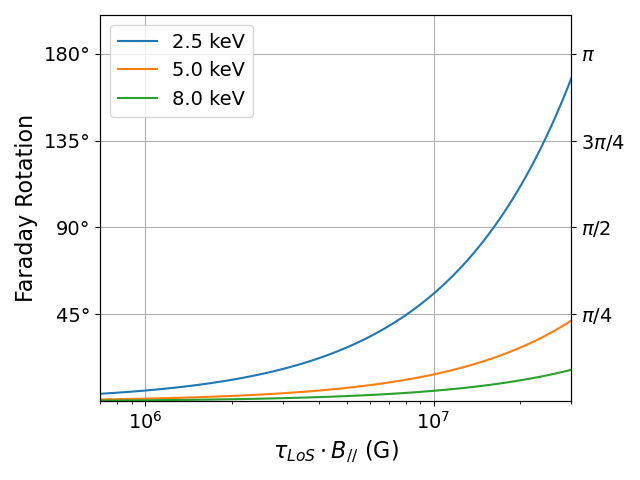}
    \caption{Faraday rotation as function of the optical depth of the line of sight and strength of the magnetic field component parallel to the line of sight for different energies of the IXPE bandpass.}
    \label{fig:faradayRotationCurves}
\end{figure}

Stokes parameters give a way to add polarized beams linearly. These are defined for linearly polarized waves as $I$, the squared amplitude of the wave, $Q$, the difference in intensities between vertical and horizontal directions, and $U$, the difference in intensities between diagonal directions of $\pi/4$ and $3\pi/4$ ($V=0$ for linearly polarized waves):
\begin{align*}
I&=\langle E_x^2\rangle + \langle E_y^2\rangle \\
Q&=\langle{E_x^2}\rangle - \langle{E_y^2}\rangle \\
U&=2\langle E_xE_y\rangle
\end{align*}
These definitions are for a single linearly polarized photon, 
so $I^2=Q^2+U^2$.

In the case of a partially polarized ($p\leq 1$) beam of photons, summing over multiple photons gives $p^2\times I^2=(Q^2+U^2)$. The Stokes parameters for a linearly polarized beam of photons can be expressed as
\begin{align*}
Q&= I\,p\,cos(2\Phi)  \\
U&= I\,p\,sin(2\Phi)
\end{align*}
Thus the polarization fraction and polarization angle 
can be recovered from the Stokes parameters 
\begin{align*}
p&= \frac{(Q^2+U^2)^{1/2}}{I} \\
\Phi&= \frac{1}{2}\arctan(U/Q)
\end{align*}
but these more intuitive quantities are not linear whereas the Stokes parameters are.

\subsection{Faraday Rotation}

\begin{figure*}
    \centering  \includegraphics[width=\textwidth,clip]{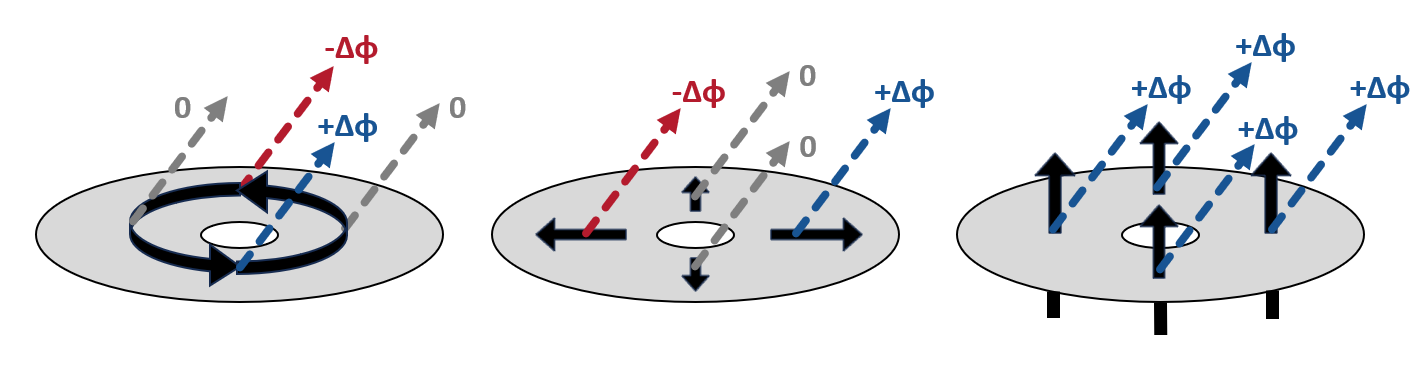}
    \caption{Faraday rotation for azimuthal (left), radial (middle) and vertical (right) magnetic field configurations. For azimuthal and radial fields there is circular symmetry so there is rotation of different sign. In the vertical field there is constant rotation across the flow. }
    \label{fig:pol_geometry}
\end{figure*}

Any plane linearly polarized wave can be split into two equal components, a left circular polarization and right circular polarization. Faraday rotation occurs where these beams travel through a magnetised plasma. The two circularly polarized directions interact differently with the magnetic field in the presence of electrons, giving a different speed of light in the magnetised plasma for left and right circular beams\footnote{see e.g. the derivation at \url{https://www.physics.rutgers.edu/~eandrei/389/Faraday_rotation.pdf}}.

Different propagation speed means that when the beams emerge from the plasma then the polarization plane is shifted by an amount (in cgs units) of
$$\Delta\Phi = \frac{e^3}{2\pi\,m_e^2\,c^4} \lambda^2 \int_{LoS} n_e(s) \, B_{||}(s) ds\ \ \ {\rm radians}$$
where $e$ is the elemental electric charge, $c$ the speed of light in vacuum, $m_e$ the mass of the electron, $\lambda$ the wavelength of the light crossing the plasma, and we integrate the product of the local electron density $n_e(s)$ 
and aligned magnetic field along the line of sight $B_{||}(s)$. We recast this from density to optical depth per unit of length $\tau_l(s)=\sigma_T \, n_e(s)$ with $\sigma_T$ the Thomson scattering cross section, such as the total optical depth along the line of sight is $\tau_{LoS}=\int_{LoS}\tau_l(s) ds$:
\begin{equation} \label{eq:FaradayRotation_integral}
    \Delta\Phi = \frac{e^3}{2\pi\,m_e^2\,c^4\,\sigma_T} \lambda^2 \int_{LoS} \tau_l(s) \, B_{||}(s) \, ds\ \ \ {\rm radians}
\end{equation}
Rewriting this in more useful units for X-ray astronomy and doing a density weighted average field along the line of sight through the plasma gives:
\begin{equation} 
    \Delta\Phi = 5.6 \ \tau_{LoS} \bigg(\frac{B_{||}}{10^6~G}\bigg) \bigg(\frac{\lambda}{5\textup{\AA}} \bigg)^2 \ \ \ {\rm degrees}
\end{equation}
\begin{equation} \label{eq:FaradayRotation}
    \Delta\Phi = 5.5 \ \tau_{LoS} \bigg(\frac{B_{||}}{10^6~G}\bigg) \bigg(\frac{E}{2.5\textup{keV}} \bigg)^{-2} \ \ \ {\rm degrees}
\end{equation}
Fig. \ref{fig:faradayRotationCurves} represents the Faraday rotation expected from Eq. \ref{eq:FaradayRotation} for different energies of the IXPE energy range. A shift $\Delta\Phi$ of $90^\circ$ means that the plane of polarization is shifted to the orthoganal direction, completely swapping the polarization. Such a rotation is equivalent to swapping the polarization from Stokes Q to -Q or from Stokes U to -U. 

Stellar mass black hole binary accretion flows have fields on order  $10^7$~G for equipartition of magnetic and gas pressure in the X-ray hot flow, or  $10^8$~G for equipartition of magnetic and ram pressure in the jet emitting flow. 
Eq. \ref{eq:FaradayRotation} 
shows that the IXPE energy bandpass (2-8~keV = 6.2-1.5 \AA ) is perfectly matched to use Faraday rotation to constrain such magnetic field in stellar mass black holes accretion flows.

\section{Effect on Observed Polarization}

In this section, we focus on computing the effects of Faraday rotation. We assume that this flow has constant intrinsic polarization fraction $p=0.05$ at all radii and azimuthal angles, with direction perpendicular to the disc (parallel to the jet axis). This translates
to Stokes parameters of $Q=0.05$, and $U=0$. We then compute how this intrinsic polarisation is modified by each field configuration. 

We assume an
inclination angle of the line of sight angle of i=30$^\circ$\footnote{Other inclination angles would only result in a factor a few difference for the limit of magnetic field strength.}.
Unless otherwise stated, we present the results for photons with energy of 2.5 keV ($\sim5$\AA).

We assume we see down to a phosphere at $\tau_{LoS}=1$ of
the X-ray emitting accretion flow. This should be the case for both bright luminous low/hard state where the total $\tau_{LoS}$ is of order 1-3 and the high/soft state where the optical depth is much larger. This means we compute the Faraday rotation expected for the photons that are escaping the flow along the line of sight and reaching the observer after their last scattering.

We assume a radial surface luminosity profile following a Novikov-Thorne emissivity profile (\citealt{NovikovThorne1973}): $F_{NT}\propto r^{-3} (1-(r_{isco}/r)^{1/2})$, where $r_{isco}$ is the radius of the innermost stable circular orbit (ISCO) and $r$ is the radial coordinate in the accretion flow. 
This is clearly appropriate for the disk dominated high/soft state but must also be similar in the bright low/hard states as seen by the continunous change in luminosity between these states \citep{Zdziarski2002, Eckersall2015} and as expected from hot flow models \citep{Xie2012}. 

We first consider the constraints for homogeneous large scale magnetic field configuration, using this to build intuition for more complex field geometries. 
In Fig. \ref{fig:pol_geometry}, we represent the three explored configurations: azimuthal, radial and vertical large scale magnetic fields. To simplify the explanation in the next paragraph, we define 4 cardinal positions from the point of view of the observer: front and back, left and right.

\begin{figure*}

    \centering
    \includegraphics[width=0.8\textwidth,trim={0cm 0cm 0cm 0cm},clip]{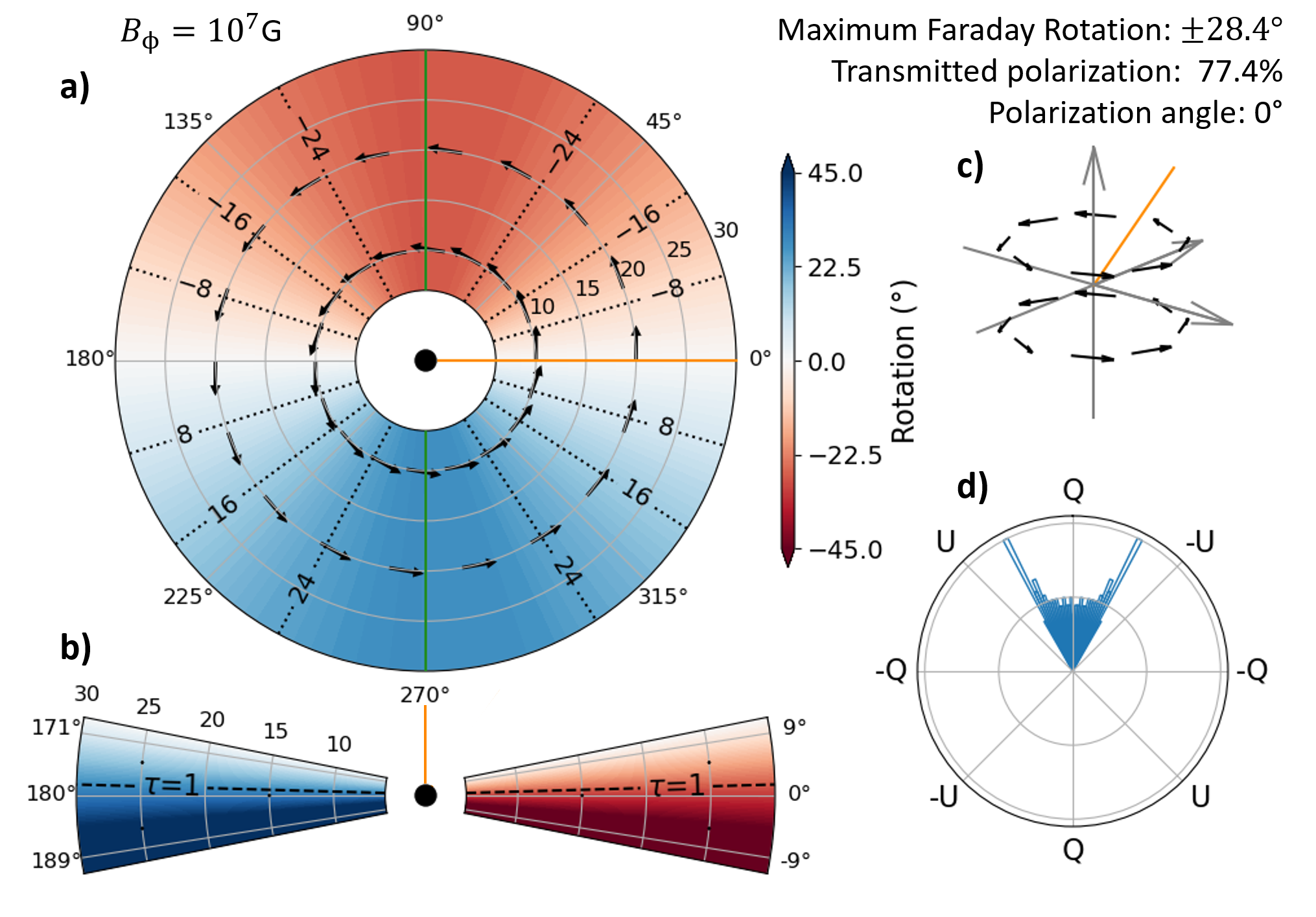} 
    \caption{a) The colormap and contours shows the Faraday rotation for a depth $\tau_{LoS}=1$ over the entire disc given a large scale azimuthal magnetic field $B_\phi=10^7$ G. The arrows show the direction of the magnetic field. The orange line shows the direction of the line of sight, fixed at an inclination angle of 30$^\circ$.  b) Cross section of the accretion flow taken at the azimuth marked in green line in a). The colormap of b) shows the Faraday rotation for escaping photons emitted at different depth inside the flow. c) shows a 3D view of the magnetic field structure and of the line of sight direction. d) shows the surface luminosity weighted distribution of the Stokes parameter of all regions of the disc shown in a). }
    \label{fig:Bphi_constant}

\end{figure*}

\begin{figure}
    \centering
    \includegraphics[width=0.45\textwidth,trim={0cm 0cm 15.2cm 0cm},clip]{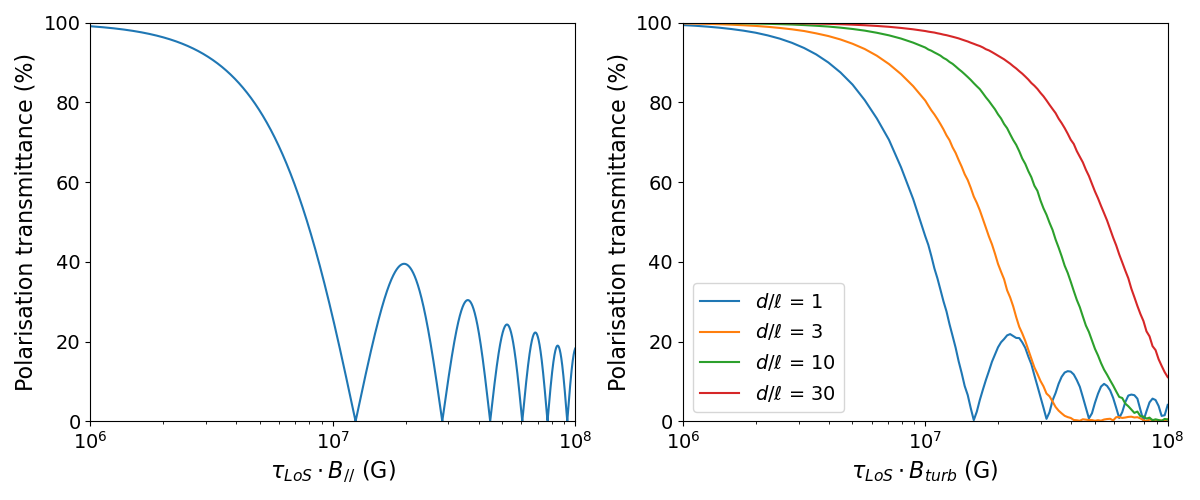} 
    \caption{ The depolarization caused by Faraday rotation for a homogeneous ordered field in either $B_\phi$ or $B_r$. }
    \label{fig:transmittance_BrBphi}

\end{figure}

\subsection{Azimuthal, $B_\phi$} \label{sec:azimuthal}

The left hand panel of Fig.~\ref{fig:pol_geometry} shows a purely azimuthal field in the X-ray hot flow. The field is perpendicular to the line of sight for the front and back of the disc, so there is no Faraday rotation. At the left and right nodal points relative to the observer azimuth, the magnetic field is partially aligned with the line of sight but it reverses on opposite sides. These left and right nodal points give equal but opposite Faraday rotation angles. This spatial incoherence of the Faraday rotation angle introduces depolarization of the disc-scale unresolved photon beam. For $\Delta\Phi=\pm 45^\circ$, the range of rotation is then $90^\circ$ large between the left and right sections of the flow, effectively depolarizing the flux from half of the disc. This will occur at a magnetic field strength $B_{||}= \sin(i) B_\phi\gtrsim 8\times 10^6$~G for photons at 2.5~keV (see Fig. \ref{fig:faradayRotationCurves}).

\begin{figure*}

    \centering
    \includegraphics[width=0.8\textwidth,trim={0cm 0.1cm 0cm 0cm},clip]{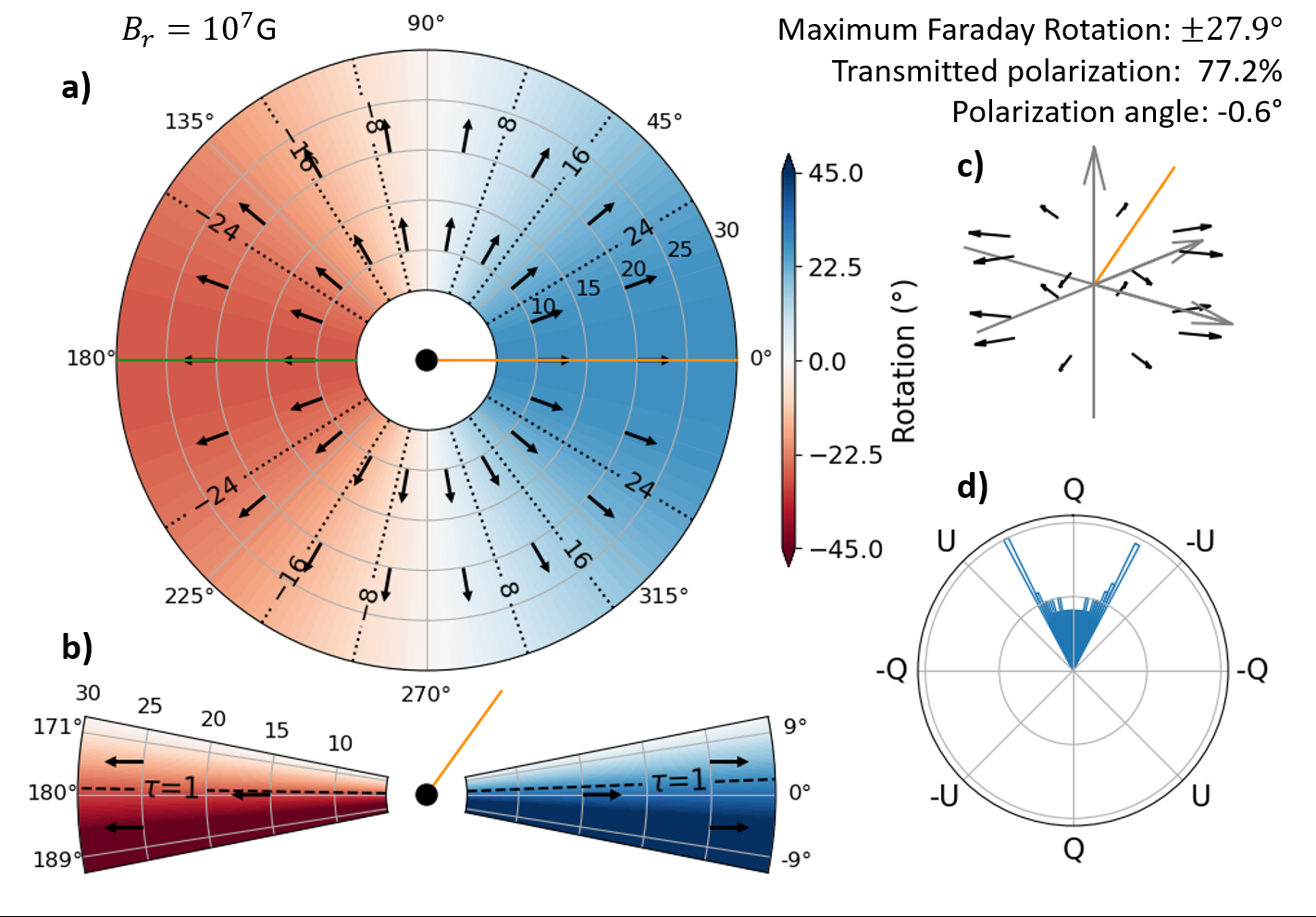} 
    \caption{Same as Fig.\ref{fig:Bphi_constant} but for constant radial field. }
    \label{fig:Br_constant}

\end{figure*}

We follow Eq. \ref{eq:FaradayRotation_integral} and integrate the optical depth and parallel magnetic field over the line of sight until the depth $\tau_{LoS}=1$. Fig. \ref{fig:Bphi_constant} represent the Faraday rotation expected for the homogeneous azimuthal configuration assuming a magnetic field strength of $10^7$ G. Panel a) shows a view from the top of the X-ray flow. The colormap represent the Faraday rotation expected at depth $\tau_{LoS}$=1 for the photons escaping along the line of sight direction (represented as an orange line stretching from the black hole). The maximal rotation is observed on the left and right nodal points, where the parallel component to the line of sight is maximal. For a $10^7$ G azimuthal magnetic field strength, the maximum Faraday rotation is $\pm28.4^\circ$. Differences to the predictions of Fig. \ref{fig:faradayRotationCurves} are due to the projection of the azimuthal magnetic field on the line of sight ($sin(30^\circ)=0.5$). Panel b) shows a cross section of the flow taken at the azimuth marked as a green line in panel a) (90$^\circ$-270$^\circ$). The colormap  shows the Faraday rotation for escaping photons depending on the depth of its emission in the flow.  The dashed line represent the $\tau_{LoS}$=1 depth where the Faraday rotation is measured. In panel c), we represent a 3D view of the geometry of the azimuthal magnetic field lines and the direction of the line of sight (in orange).  In panel d), we represent the distribution of the luminosity weighted Stokes parameters for the different photon beams coming from the surface of the flow. The more spatially incoherent the Faraday rotation is over the flow, the broader the  distribution in the Stokes parameter space becomes, reducing the polarization fraction of the total unresolved beam. For this particular configuration, assuming an inclination angle of 30$^\circ$, and a $10^7$ G purely azimuthal magnetic field structure, only 77.4\% of the initial polarization fraction is transmitted.

Fig. \ref{fig:transmittance_BrBphi} shows the polarization transmittance for the azimuthal configuration as function of the strength of the parallel magnetic field component. 
This shows the depolarization due to the spatial incoherence introduced by Faraday rotation. The behaviour is not intuitive due to the non linearity of the problem and the phase-wrapping that intervenes at high magnetic field strength. Interestingly, a parallel field of $\sim 1.3\times10^7$ G will result in a completly unpolarized beam. The best way to understand this is through the distribution of the Stokes parameter over the entire flow. To get an unpolarized beam,  both the Q and U Stokes parameters need to cancel out. This happens for a parallel magnetic strength of $B_{||}\sim 1.3\times10^7$ G. Above this value, the Stokes parameter starts to phase-wrap and the corresponding Stokes distribution will get larger and flatter, resulting in a smaller polarization transmittance envelope.

\begin{figure*}

    \centering
    \includegraphics[width=0.8\textwidth,trim={0cm 0.1cm 0cm 0cm},clip]{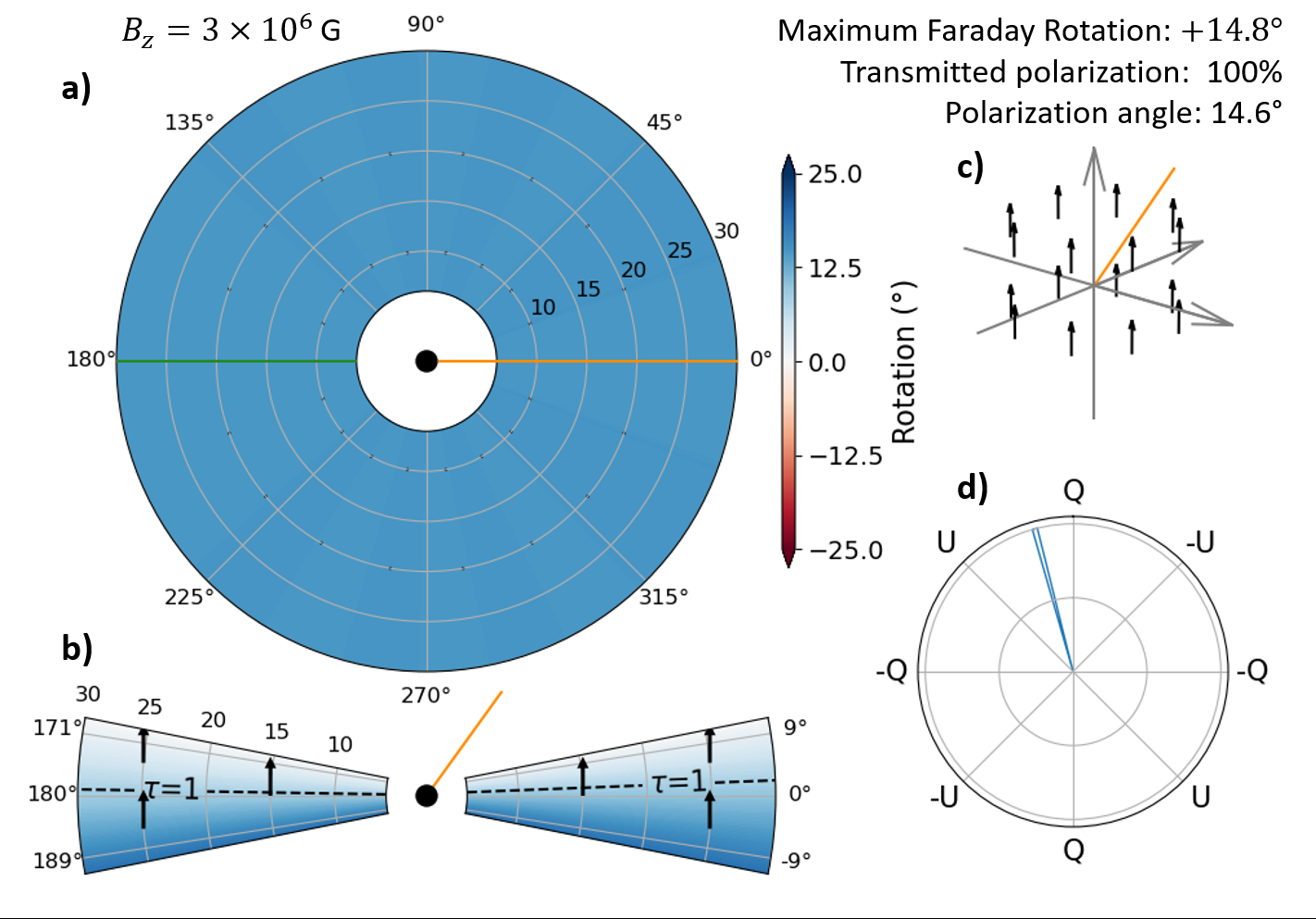} 
    \caption{As in Fig.\ref{fig:Bphi_constant} but for constant vertical field}

    \label{fig:Bz_constant}

\end{figure*}

\subsection{Radial, $B_r$}

The middle panel of Fig.~\ref{fig:pol_geometry} shows a purely radial field. This geometry means that the left and right nodal points have no line of sight component, but the front and back of the flow have equal and oppposite Faraday rotation angles. Again, for $\Delta\Phi=\pm 45^\circ$, effectively depolarizing the flux from half of the disc. 

In Fig. \ref{fig:Br_constant}, we plot the Faraday rotation obtained from the integration of Eq. \ref{eq:FaradayRotation_integral} along the line of sight for the entire flow given a radial magnetic field strength of $10^7$ G. The panels a) to d) are the same as in Fig. \ref{fig:Bphi_constant} and described in Sect. \ref{sec:azimuthal}. Except for the cross section shown in panel b) which is now taken along the x-axis  (0$^\circ$-180$^\circ$). Compared to the azimuthal geometry case, the Faraday rotation map in panel a) has rotated by $90^\circ$ azimuth. The maximum of the rotation are now at the front and back, whereas the left and right nodal points have no Faraday rotation.

The transmittance curve for the radial geometry is then essentially the same as the transmittance curve for the azimuthal geometry plotted in Fig. \ref{fig:transmittance_BrBphi}. The small differences observed between the radial and azimuthal cases results from second order geometry effects of the line of sight inclination with the disc height scale. 

\subsection{Vertical, $B_z$}

The right hand panel of Fig.~\ref{fig:pol_geometry} shows a purely vertical field, the most interesting geometry for jet launching. The field always makes the same angle to the line of sight, so there is a constant Faraday rotation angle across the disc. Since no spatial incoherence of the polarization angle is introduced by the vertical geometry, no depolarization is expected. 

In Fig. \ref{fig:Bz_constant}, we plot the Faraday rotation obtained from the integration of Eq. \ref{eq:FaradayRotation_integral} along the line of sight for the entire flow given a vertical magnetic field strength of $3\times10^6$ G. The panels a) to d) are the same as in Fig. \ref{fig:Bphi_constant} and described in Sect. \ref{sec:azimuthal}, except that the cross section shown in panel b) is now taken along the x-axis  (0$^\circ$-180$^\circ$). A constant vertical magnetic field geometry does not introduce any incoherence in the polarization angle and thus, the initial polarization fraction is conserved. However, the polarization angle has now entirely rotated by 14.6$^\circ$. This is for an emission at 2.5 keV ($\sim5$\AA). Because of the dependancy of Faraday rotation with the wavelength, at 8 keV ($\sim1.55$\AA), the expected rotation of the polarization angle will only be of about 1.5$^\circ$. Stronger magnetic fields will only result in a larger difference within the IXPE bandwidth. This can directly be compared to the data obtained for the polarization angle as function of energy, bringing strong constraints on the maximum allowed vertical magnetic field component inside the flow. One can refer to Fig. \ref{fig:faradayRotationCurves} to see how much the polarization angle will rotate because of the vertical magnetic field for a few energies.

\section{Jet launching in Cyg X-1}

We showed that azimuthal and radial   magnetic field components introduce incoherence in the polarization angle over the disc, reducing the polarization fraction of the unresolved beam that will be observed by IXPE. By contrast, the vertical magnetic field component rotates the entire polarization angle of the flow. 
IXPE can measure both polarization fraction and angle as a function of energy, which means it can set 
constraints on the values of magnetic field components inside of the photosphere of the accretion flow. 

\begin{figure}
    \centering
    \hspace{-1cm}
    \includegraphics[angle=90,width=0.5\textwidth,trim={7cm 2cm 4.5cm 15cm},clip]{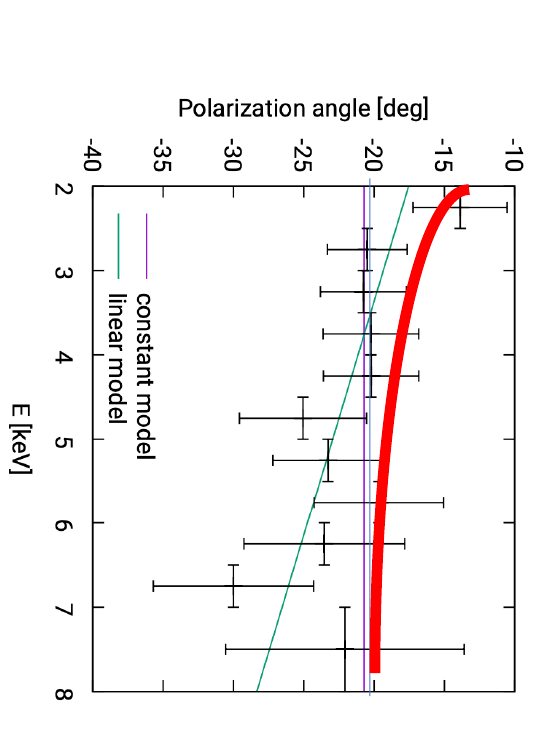}
    \caption{Cyg X-1 polarization angle as a function of energy (black data) showing its general alignment with the radio jet axis (magenta line), though there is a marginally significant linear trend (green line). The red line shows the predicted Faraday rotation angle from a homogeneous field of $B_z=10^6$~G. Adapted from \cite{Krawczynski2022}}
    \label{fig:cygx1_pa}
\end{figure}

The polarization angle seen in Cyg X-1 is aligned on average with the radio jet. Assuming this is the intrinsic polarization direction, we compare to the data in Fig.\ref{fig:cygx1_pa}, where the magenta line indicates the jet direction. The red line shows the effect of a homogeneous vertical field of $B_z = 10^6$~G threading the X-ray emitting plasma. This is clearly consistent with the data, but Faraday rotation goes linearly with field strength so even $B_z = 2-3\times 10^6$~G is strongly inconsistent. 
This gives an upper limit to the vertical field strength which is consistent with the data for an initial polarization angle intrinsically aligned with the jet. Any other initial polarization angle appears fine tuned.

Given the polarization transmission plotted in Fig. \ref{fig:transmittance_BrBphi}, any strong azimuthal  or radial magnetic field component will reduce the polarization of the accretion flow by a large amount ($\geq 50\%$) for values above $B_{r,\phi} \sin(i) = 8\times10^6$ G. It is already difficult to explain the $\sim$5\% polarization fraction observed for Cyg X-1 when not taking into account any Faraday rotation effects. However, if azimuthal or radial magnetic fields stronger than $10^7$~G are present, this would require 
even larger 
intrinsic polarization in the flow so that Faraday rotation depolarization still gives a value of $p=0.05$. 
As such, we believe no strong Faraday rotation is present, putting an upper limit on the magnetic field strength inside the flow. 

The large scale magnetic field upper limits inferred from Faraday rotation are very low field strengths compared to the ones proposed by many models (see discussion in  Sec. \ref{sec:low-hard}). 

\section{More complex field geometries}

\subsection{Radial stratification of the magnetic field}

Most models have $B_i\propto r^{-\alpha}$, where $i$ is one of directions $r,z,\phi$, so the amount of Faraday rotation will change radially across the X-ray emission region. The surface emissivity, $\epsilon(r)$, also has radial dependence. 

What sets the amount of Faraday rotation/depolarization is the
emissivity weighted mean field\footnote{It should be emissivity {\it and} optical depth weighted mean field however we assume that the optical depth is constant with radius.} $\langle B_i \rangle$ over the surface of the disc:
\begin{equation} \label{Eq:rad_stratification}
    \langle B_i\rangle = \frac{\int_{r_{isco}}^{r_{out}} \epsilon(r) 2\pi r B_i(r) dr }
{\int_{r_{isco}}^{r_{out}} \epsilon(r) 2\pi r  dr}
\end{equation}

Constraints on the vertical field give that the Faraday rotation needs to be $\lesssim 8^\circ$ at $2.5$~keV, so this is the same limit as before but now on the emissivity weighted mean field $\langle B_z\rangle \lesssim 2\times10^6$~G. 

For the specific case of a Novikov-Thorne emissivity profile for a flow extending from $6$ to $30R_g$ and with $\alpha=5/4$ (as required in the analytic ADAF hot flow models \citealt{Narayan1995} and jet emitting disk models \citealt{Marcel2018}), this gives a magnetic field strength at the ISCO of $B_z(r_{isco})\sim 6\times 10^6$~G. Steeper radial dependence of the field allows larger $B_z(r_{isco})$.

Similarly for the radial and azimuthal field components, the limit is now on the emissivity weighted mean field $\langle B_{r,\phi} \rangle \sin(i) \lesssim 8\times 10^6$~G. Again, this allows higher field on the inner edge of the flow for standard Novikov-Thorone emissivity ($B_{r,\phi}(r_{isco})\sim4\times 10^7$~G), but this still sets a stringent upper limit to the magnetic pressure inside the flow. 

In the previous computation and figures we assumed a non rotating black hole with ISCO at 6 $R_g$. However, if the black hole is spinning, this value can be lower. Keeping the same magnetic field profile at the outer edge results in stronger magnetic field strength at the inner edge since the disk now extends closer to the black hole. This results in stronger Faraday depolarizataion/rotation effects for the X-ray photons. We note that in the case of a misaligned disk, such effect of the spin is not expected as the inner edge of the flow remains almost constant with the spin \citep{Fragile2009}. However the magentic field structure in this case would be more complex, and each simulation should be tested individually.

Multi-temperature corona models can have the 2 keV and 8 keV emission coming from different radii inside the flow. Combined with the radial dependence of the magnetic field, this could also reduce the expected difference in polarization fraction or angle within the IXPE bandpass. 

\subsection{Vertical stratification of the magnetic field}

One way to hide the effect of the magnetic field is to bury it below the photosphere. The field can then be dynamically important in the bulk of the flow, but small 
enough in the region above the last scattering surface to not cause significant depolarization by Faraday rotation.

This requires a model of the vertical density and magnetic field structure in order to evaluate its effect, but in general this will not change the constraints for low/hard states as the flow is not very optically thick, so there is nowhere to hide the field. 

\subsection{Turbulent fields}

The MRI dynamo gives turbulent fields, 
and these (surprisingly) give much less depolarization than large scale ordered fields. 
Defining the turbulence length scale $\ell$ and the mean free path $d$, we can describe the $\tau_{LoS}=1$ photosphere with a typical number of cells along the line of sight of $N=d/\ell$. Each of these has a random magentic field direction, but only contributes over an optical depth $d\tau=1/N$ of the total path. Summing over the line of sight then gives a result which strongly depends on the size scale of the turbulence. If the turbulence is large, so there is only 1-2 cells in the photosphere, then the photons only see an ordered large scale magnetic field structure, similar to the cases discussed before. When the number of cells increases, there are more field reversals, and each contributes for a shorter amount to the total optical depth path length. 

This can be described by a random walk algorithm where at each cell or 'step', the polarization angle rotates by an amount $\Delta\Phi$ which depends both on the local turbulent magnetic field stength and direction and the optical depth contribution of the cell. In other words, it is a random walk where the maximum length of the step depends on the number of steps. For a large number of steps, one can show that such a random walk tends to a Gaussian distribution peaking around the initial position, so on average the Faraday rotation will be small. 

\begin{figure}
    \centering
    \includegraphics[angle=0,width=0.5\textwidth,trim={0cm 0cm 0cm 0cm},clip]{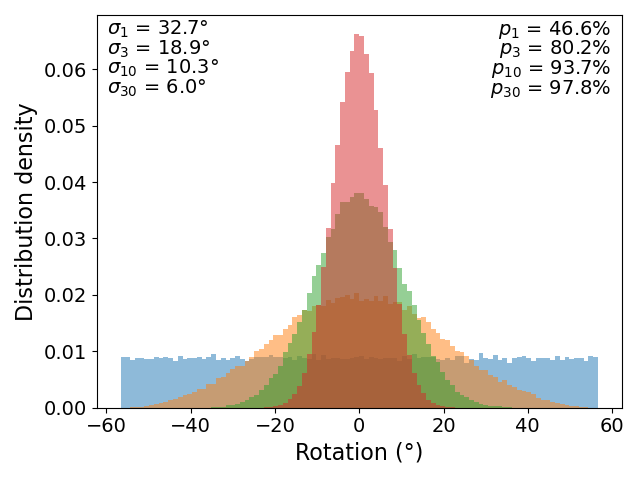}
    \caption{Distribution of Faraday rotation for 10$^5$ photons crossing the $\tau_{LoS}=1$ photosphere with a turbulent magnetic field. The results depends on the typical scale of the turbulence $d$ compared to the length of the photosphere $\ell$.  In blue, orange, green and red are represented the cases of N = d/$\ell$ = 1, 3, 10 and 30 respectively. The larger the number of cells in the photosphere, the smaller the effects of Faraday rotation become.   }
    \label{fig:Bturb_Histogram}
\end{figure}

\begin{figure}
    \centering
    \includegraphics[width=0.45\textwidth,trim={15.2cm 0cm 0cm 0cm},clip]{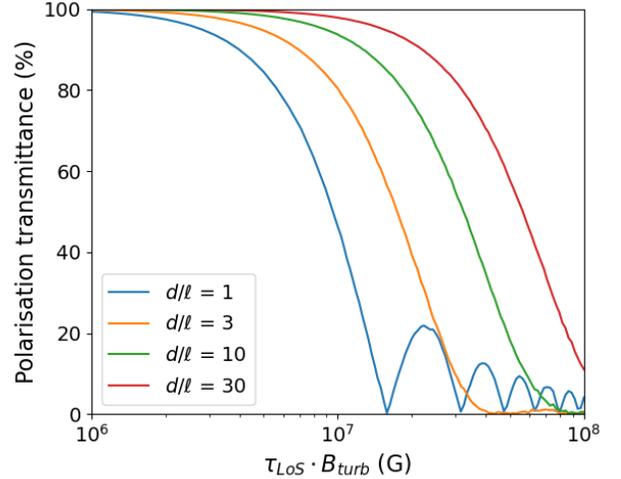} 
    \caption{Depolarization caused by Faraday rotation for a turbulent field with different length scales of turbulence $\ell$ relative to the mean free path $d$. The more distinct cells, the more random fields along the line of sight, but each one is over a smaller length scale so counterintuitively the depolarization is reduced by multiple turbulent cells. }
    \label{fig:transmittance_Bturb}

\end{figure}

In Fig. \ref{fig:Bturb_Histogram}, we plot the distribution of Faraday rotation  for different number of cells in a line of sight of the $\tau_{LoS}=1$ photosphere. In each case, we draw $10^5$ photons with the same initial polarization angle. The magnetic field strength of the turbulence is fixed to $10^7$ G so that the maximum Faraday rotation possible, when all the field is aligned with the line of sight, is  $56^\circ$. For each photon and in each cell, we draw a random direction  of the magnetic field with a 3D isotropic probability. As such the size of the step in each cell is stretching randomly between [- 56/N $^\circ$ ; + 56/N $^\circ$] depending on the projection of the magnetic field direction on the line of sight.
If there is a single turbulent cell, this results in a flat distribution stretching from -56$^\circ$ to +56$^\circ$. When the number of cell increases, the distribution tends toward a gaussian shape centered around 0 with decreasing dispersion. As such if the number of cell is larger than 3, the amount of incoherence introduced by the Faraday rotation will be small enough to not lose too much polarization fraction.

In Fig. \ref{fig:transmittance_Bturb}, we plot the transmittance curve for a single line of sight as function of the line of sight optical depth and turbulent magnetic field strength for different number of cells. For N=1, the transmittance curve is similar to the one obtained for the ordered large scale radial and azimuthal magnetic field geometry for the entire disc. When the number of cells increases, stronger magnetic fields are required to reduce the polarization fraction. As such, where one would intuitively expect that a small scale turbulent field would completly unpolarize the beam, we see that the smaller the tubulence is compared to the photosphere, the easier it is to conserve the initial polarization. This concerns a single line of sight. But one should expect the beams coming from other regions of the turbulent disc to have the same distribution and so we can scale up these distributions to the entire flow.

\section{Models of jet launching in the low/hard state} \label{sec:low-hard}

There is a small scale turbulent magnetic dynamo produced by the magneto-rotational instability (MRI: \citealt{Balbus1991}) which amplifies any weak magnetic field in the accretion flow. 
Decades of work on the MRI have shown that the instability grows linearly, then saturates in the non-linear regime, giving a turbulent 
magnetic field structure with well defined average properties which transport angular momentum radially outwards, allowing material to accrete radially inwards. 

The discovery of the MRI held out the hope that the magnetic field configuration and consequent jet launching would emerge {\it ab initio} from these models. 
However, it is now clear that the properties of the dynamo depend on the net magnetic flux imposed as part of the initial conditions. Simulations with no net flux e.g. initialised as a single loop of weak magnetic field inside a plasma torus evolve to a steady state where the bulk of the flow is dominated by turbulent field, but with an inner
large scale $B_z$ component which threads both the inner parts of flow (producing the funnel wall accretion powered jet) and the black hole horizon (producing a spin powered jet). However this large scale $B_z$ is rather weak, and shows field reversals over time. These  zero net flux flows are termed Standard and Normal Evolution (SANE).
Angular momentum transport is mostly via the turbulent dynamo, giving an effective $\alpha$ viscosity 
which is tied to the ratio of gas pressure to magnetic pressure in the flow, $\beta=P_{gas}/P_B$ (\citealt{Begelman2015,Salvesen2016,Begelman2023}).
The best known analytic/numerical approximations to these flows are the Advection Dominated Accretion flows \citep{Narayan1995}.
For Cyg X-1, with $\dot{m}\sim 0.02$ and a borderline weakly magnetised plasma has $\beta\sim 10$ (which corresponds to 
\cite{Narayan1995} $\beta$ parameter of $\beta/(\beta+1)=0.91$) giving $B\sim 2\times 10^7 r^{-5/4}~G$ for a $20~M_\odot$ black hole and $\alpha=0.1$. For $r=10$, around where the emissivity peaks, this gives $B\sim 10^6~G$.
This is clearly consistent with the constraints from Faraday rotation, especially for turbulent fields. 

Apart from zero net flux (SANE models), the only other natural configuration appears to be the maximum magnetic flux which can be held onto the black hole by the accretion flow i.e. where the magnetic pressure is of order the ram pressure of the infalling material. These flows were originally discussed in 1D, so were called magnetically arrested discs (MAD), as in 1D the flow is completely halted when  $P_B\sim P_{\rm ram}$. However, accretion does actually continue in 3D as there are interchange instabilies which allow blobs of matter to accrete. MAD flows produce powerful jets and have more efficient angular momentum transport from the torque exterted by the large scale fields on the inflowing matter.
For Cyg X-1 low/hard state, the mass accretion rate of $\dot{m}\sim 0.02$ corresponds to $\dot{M}\sim 5\times 10^{17}$~g/s. This in infalling with velocity $\sim c$ at the horizon so its ram pressure is of order $\dot(M) c /(4\pi R_g^2) $, so $B^2\sim 2\dot{M} c/R_g^2$ giving $B_z\sim 6\times 10^7$~G. 

An alternative, more formal way to define a MAD flow is via the dimensionless magnetic flux on the horizon 
\begin{equation}
    \phi_{\rm BH} = 50 = \frac{<B^r 2\pi R_g^2>}{<\dot{M} R_g^2 c>^{1/2}} =  \frac{<B^r 2\pi R_g>}{<\dot{M} c>^{1/2}}
    \label{}
\end{equation}
where $B^r$ denotes the radial component of the field in spherical polar coordinates.
\begin{equation}
    \frac{50}{2\pi R_g} {<\dot{M} c>^{1/2}} = B^r \sim 4\times 10^8~{\rm G}
    \label{}
\end{equation}

Unsurprisingly, the MAD flows have stronger fields than the SANE and these are far above the limits on large scale fields derived above from Faraday rotation. 
However, the MAD flows typically put this strong field onto the horizon itself, whereas the Faraday rotation limits only apply to the field inside the X-ray emitting flow, so these require a specific model of the field and flow densities. 

A good semi-analytical approximation to MAD flows with large scale poloidal fields are the Jet Emitting Disc (JED) models (\citealt{Ferreira97}). These assume the jet is launched from the flow itself through the BP process, (\citealt{BP1982}) and its feedback magnetic torque extracts angular momentum vertically from the accretion flow. This results in a supersonic accreting flow which becomes optically thin and hot (\citealt{Marcel2018} and references therein). JED are generally characterized through the magnetization parameter $\mu=2P_{B_z}/P_{tot}$ taken at the midplane\footnote{Note that $\mu$ is not simply the inverse of the plasma $\beta$ as it depends on the vertical magnetic field components and not all components ($B_{r,\Phi,turb}$) as well as the total (sum of gas and radiative) pressure.}. The range of values of $\mu$ are within [0.1-1]. Assuming the values for Cyg X-1 (obtained from fitting the observation, W. Zhang in preparation), the expected magnetic field strength is $B_z = 7.8\times10^7~\big(\frac{\mu}{1}\big)^{1/2}~r^{-5/4}$ G (\citealt{Marcel2018}). This shows that only weakly magnetized discs with $\mu\lesssim0.1$ can have low enough magnetic fields to avoid Faraday rotation effects at 10-20 $R_G$, where most of the 2 to 8 keV emission is produced. Lower vertical fields may not be strong enough to give sufficient torque on the disc to make the transition to a supersonic inflow (e.g. \citealt{Jacquemin2019}). 

\subsection{Consequences on effective viscosity}

Simulations show that the effective viscosity and plasma $\beta=P_{\rm gas}/P_{\rm B}$ parameter are linked together, with $\alpha=11 \beta $ \citep{Salvesen2016,Mishra2020}. This implies that large effective viscosity requires large magnetic fields, as large scale magnetic torques efficiently extract angular momentum. Large effective viscosity is required by observations of black hole binary outbursts \citep{King2007,Kawamura2022, Tetarenko2020}. The limits on Faraday rotation then require that these fields are turbulent rather than coherent.

\subsection{Examples of possible field configurations}

If there is a powerful jet then this requires 
substantial vertical magnetic field. The 
discussion above make it clear that  
there are strong constraints on the field strength which threads the X-ray emitting hot plasma, challenging all BP models. Instead, some MAD flows have the strong vertical fields mostly outside of the plasma, confined between the ISCO and the event horizon (e.g. \citealt{Liska2024}, see the youtube links from that paper for the plasma $\beta$ plots). Strong separation of the field and X-ray emitting plasma will not produce Faraday rotation, so these configurations, where the jet is powered by the BZ effect, can 
match the observational constraints. 

Nonetheless, there must be magnetic fields in the X-ray emitting plasma as there must be some form of angular momentum transport. Matching all the constraints then gives a potential solution where the flow is dominated by the small scale, turbulent MRI 
dynamo, with strong (MAD) field only on the event horizon to launch a powerful jet. However, not all MAD flows show this configuration: e.g. the RADPOL simulations of \citet{Liska2022} is MAD, but has substantial (magnetisation of 10) ordered poloidal field inside the X-ray emitting flow. which is clearly challenged by the observed polarization in the low/hard state.

Even this potential solution may be open to observational challenge from polarisation as magnetic field can also rotate the plane of polarisation of light from vacuum birefringence. This becomes important when 
$B_\perp \gtrsim 10^8 
(d/1.5\times 10^6~\rm{cm})^{-1/2} 
(E/2.5~keV)^{-1/2}$~G 
(rescaled from 
\citealt{Krawczynski2021} for 
$\Delta\theta=5.7^\circ$ rather than 
1~radian as well as changing typical 
distance to $R_g$ for a $10M_\odot$
and energy to 2.5~keV). These are the 
typical fields strengths around the 
horizon for MAD flows (see also \citealt{Caiazzo2018}), though the 
observational consequences depend on 
quantifying how much of the X-ray flux 
crosses this high field region. 

Instead, all the constraints are trivially circumvented if the jet is not powerful. This is the case for models where the jet is seeded by photon-photon collisions 
producing electron-positron pairs \citep{Zdziarski2022,Zdziarski2024}. The much lower power of these light jets mean that these can 
be accelerated by much lower fields, so do not have significant impact on the polarisation. However, these are also well below equipartition  
so cannnot significantly affect the dynamics of the flow. 

\section{Applications to disc physics in high/soft states}

Black hole binaries make a dramatic spectral transition from the low/hard state, where the emission is dominated by Comptonisation
from hot, optically thin plasma, to a high/soft state, where the emission is dominated by an optically thick, thermal disc. The switch from optically thin to optically thick predicts that the polarization should change, even if the source geometry is a radially extended disc-like plane in both states.
This is because of the switch in seed photon direction. The seed photons have to be travelling predominantly in the plane of the hot flow in order to encounter an electrons in an optically thin source, whereas they are predominantly vertical in an optically thick source. 
This $90^\circ$ swing in seed photon direction before the last scattering predicts that there is a $90^\circ$ swing in polarization angle of the scattered photons, from being perpendicular to the disc (aligned with the jet) for the optically thin low/hard state, to being aligned with the disc (perpendicular to the jet) in the optically thick high/soft state
(see e.g. \citealt{Tomaru2024}).

There are now some IXPE observations of polarization in the high/soft state, but most of these are for sources where the jet direction is not known (LMC X-1,  LMC X-3, 4U1630-47, 4U1957+115). None of 
these have comparison data for a low/hard state, so these cannot be used to test the predicted switch in direction either. Nonetheless, these data do give some
information, especially in LMC X-3 where the spectrum is clearly disc dominated, the system parameters are well determined, and the inclination is high enough ($\sim 70^\circ$) that there should be polarization. 
The data show $p=0.032\pm 0.006$, consistent with expectations from an optically thick, geometrically thin disc \citep{Svoboda2024a}. LMC X-1 has no detected polarization, which is consistent with its lower inclination angle \citep{Podgorny2023}, 4U1957+115 has similar polarization to LMC X-3 at $\sim 0.019\pm 0.006$, but the source distance and inclination are unknown \citep{Marra2024}, while 4U~1630-47 has surprisingly high polarization of $0.08$, even for a highly inclined disc \citep{Ratheesh2024}.

There are two exceptions, where the same source is seen in both high/soft and low/hard states, and where the jet direction is known. One is Cyg X-1, where there was surprisingly no change in polarization direction between the two states, with the only difference being a small drop in polarization fraction \citep{Dovciak2023,Jana2024}. However, Cyg X-1 probably does not show a true soft state \citep{Zdziarski2024cyg,Belczynski2021}, which may make this more complex. By contrast,
Swift J1727 shows a clean high/soft state, with $p\lesssim 0.01$ \citep{Svoboda2024b} after showing 
polarization of $0.04$ in the hard (intermediate) state \citep{Ingram2023}, parallel to the jet, similar to the low/hard state of Cyg X-1. The inclination is not yet known, but seems likely to be lower than in LMC X-3 and 4U1957. 

On balance then it seems likely that the clean disc spectra seen at $\sim 60-70^\circ$ inclination show $p\sim 0.02$, as expected from electron scattering in a 
plane parallel atmosphere 
\citep{Sunyaev1985}. The lack of evidence for  strong depolarization from Faraday rotation puts limits on large scale fields in the disc photosphere. This is especially important for the toroidal field, $B_\phi$, as it is this component which often invoked to stabilise the disc against the thermal instability \citep{Blaes2006,Begelman2007,Bai2013,Begelman2024}. In order to do this, the field must be dynamically important, at or above equipartition, and the higher mass accretion rates seen in the soft states mean that this is $\geq 10^7$~G, easily above the Faraday rotation limits. 

However, the high/soft state is optically thick, so a strong toroidal field could be dominant on the equatorial plane, controlling the dynamics, but much lower in the photosphere, so not producing Faraday Rotation. This is not seen in simulations, instead the buoyancy of the field tends instead to a configuration where the photosphere is very highly magnetised compared to the midplane \citep{Begelman2024}. 
While we stress that simulations should be checked individually against the Faraday Rotation constraints, it seems most likely that strong toroidal fields can be ruled out as the origin for the observed lack of thermal instability in black hole binary discs. 

\section{Conclusions}

The energy range of IXPE is perfectly matched for testing the magnetic field configuration in X-ray binary black holes. Quite general arguments show that equipartition fields in these systems should be of order $10^{6-8}$~G. If these are ordered on large scales then they produce observable Faraday Rotation and/or depolarization in the 2-8~keV band. Since there is no evidence for this,
it seems likely that there are no equipartition strength large scale ordered fields in the black hole binary discs. 

This is completely unexpected in both the low/hard and high/soft states, where there was growing consensus that these were dominated by strong, ordered fields. For the low/hard state, the picture was that jets launching was via large scale poloidal fields from MAD or JED flows. This is still potentially possible for configurations where the strong poloidal field separates from the X-ray emitting flow, dominated by the small scale dynamo. Here there can be a powerful jet, tapping the spin energy of the black hole from the horizon threading (BZ configuration) but not all MAD flows show sufficient separation of field and flow, and even those that do may give significant rotation of polarisation from vacuum birefringence. 
Another possibility is that jets are launched from weakly magnetized discs (BP configuration) but whether these are powerful enough remains unclear. 

Conversely, in the high/soft state, the consensus had been that strong toroidal fields supressed the thermal instability. Again, this is strongly challenged by the observed polarization. 

Fundamentally, polarisation data from IXPE makes the invisible visible, allowing the specific predicted density/magnetic field configuration derived from simulations to be tested against the observations, guiding us to a better understanding of accretion physics. 

\begin{acknowledgements}
SB is an overseas researcher under the Postdoctoral Fellowship of Japan Society for the Promotion of Science (JSPS), supported by JSPS KAKENHI Grant Number JP23F23773. CD acknowledges support from STFC through grant ST/T000244/1, and 
University of Tokyo Kavli/IPMU for hospitality and support for a visit during which much of this work was produced. Kavli IPMU was established by World Premier International Research Center Initiative (WPI), MEXT, Japan.
SB and CD thank Adam Ingram and Jonathan Ferreira for extremely helpful conversations about polarization, and Matthew Liska for sharing simulation results. We thank Jeremy Heyl for bringing vacuum birefringence to our attention. 

\end{acknowledgements}

\bibliography{biblio}{}
\bibliographystyle{aasjournal}

\end{document}